\documentclass{article}
\usepackage{graphicx}
\usepackage{braket}
\usepackage{amsmath}

\begin{document}

\title{On Statistical Independence and No-Correlation for a Pair of Random Variables Taking Two Values: Classical and Quantum}
\author{Toru Ohira\thanks{The author is also affiliated with Future Value Creation Research Center, Graduate School of Informatics, Nagoya University.}\\
Graduate School of Mathematics, Nagoya University, Japan}

\maketitle

\begin{abstract}
It is well known that when a pair of random variables is statistically independent, it has no-correlation (zero covariance, $E[XY] - E[X]E[Y] = 0$), and that the converse is not true. However, if both of these random variables take only two values, no-correlation entails statistical independence. We provide here a general proof. We subsequently examine whether this equivalence property carries over to quantum mechanical systems. A counter-example is explicitly constructed to show that it does not. This observation provides yet another simple theorem separating classical and quantum theories.
\end{abstract}

\section{Introduction}

Differences and boundaries between the 
classical and quantum mechanics, as manifested by the Bell's inequalities\cite{bell,sakurai}, have been under active research investigations both
theoretically and experimentally.  In this paper, we present yet another simple theorem which separates 
classical and quantum probability theories. 

It is well known and can be simply proven that when two random variables are statistically independent,
they are not correlated. The converse is not true in general (e.g., \cite{feller}). We can have a pair of random variables which is not
correlated but not statistically independent, and such examples can be easily constructed as well. (A known exception is when their joint probability density function has the form of a bivariate normal distribution\cite{bain}.)

We, however,  first show that when both of these random variables are not continuous and take only two distinct values, statistical independence and no-correlation become equivalent in general. In other words,
the proposed theorem means that one cannot have an uncorrelated pair of random variables with two distinct values which is not statistically independent.

We, then, consider the corresponding question in quantum mechanics. The notion of statistical independence is associated with that of separability of state vectors for a two-particle quantum system. We can see that the separability entails no-correlation as in the classical probabilistic framework. However, 
the above-mentioned theorem about two-state systems does not carry over to quantum mechanics.
We show this by explicitly constructing a counter-example.

\section{Classical case}

Consider two random variables $X$ and $Y$, such that they both take only two distinct finite values $(x_1, x_2)$ and $(y_1,y_2)$. Denote the joint probability distribution for these variables as
$P(X:Y)$, and assume it is given by
\begin{equation}
p({x_i}:{y_j}) \equiv P(X={x_i}:Y={y_j}) = p_{ij}, \quad ( i,j \in \{1,2\} )
\label{jp}
\end{equation}
Then, the probability distributions $P(X)$ for $X$ and $P(Y)$ for $Y$ are simply expressed as follows.
\begin{equation}
p({x_i}) \equiv P(X={x_i}) = p_{i1} + p_{i2}, \quad
p({y_j}) \equiv P(Y={y_j}) = p_{1j} + p_{2j}.
\label{ep}
\end{equation}
By the requirement that both $X, Y$ take only two values,
\begin{equation}
p({x_1}) + p({x_2}) = p({y_1}) + p({y_2}) = 1.
\label{sp}
\end{equation}

These relations can be summarized in the following table.
\begin{figure}[htb]
\begin{center}
\includegraphics[width=0.23\columnwidth]{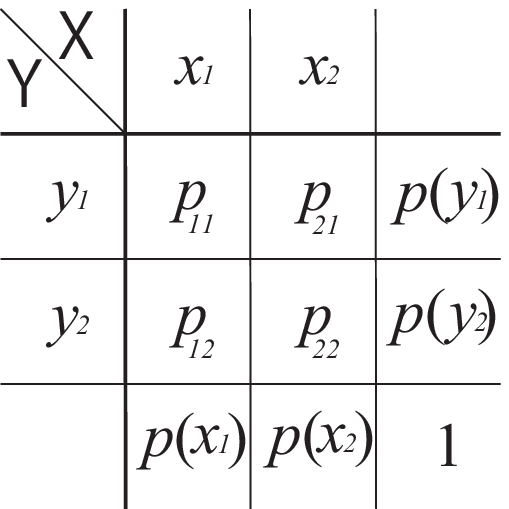}
\end{center}
\label{sind1}
\end{figure}

The statistical independence of $X$ and $Y$ is defined as 
\begin{equation}
P(X:Y)=P(X)P(Y).
\label{ind0}
\end{equation}
Also, with the definition of expectation values as 
\begin{equation}
E[X] = \sum_i p({x_i}){x_i},\quad
E[Y] = \sum_i p({y_i}){y_i},\quad
E[XY] = \sum_{i,j} p({x_i}:{y_j}){x_i}{y_j},
\label{ev}
\nonumber
\end{equation}
we define that $X$ and $Y$ are not correlated when their covariance is zero. 
\begin{equation}
Cov[X,Y] \equiv E[XY] - E[X]E[Y] = 0,
\nonumber
\end{equation}
or equivalently, 
\begin{equation}
E[XY] = E[X]E[Y].
\label{noc0}
\end{equation}

We can see from the above definition that given (\ref{ind0}), (\ref{noc0}) follows. Our main statement here is 
that the converse is true, i.e., (\ref{ind0}) and (\ref{noc0}) are equivalent when both of these random variable take two finite distinct values. In passing, we note that if either (or both) $X$ or $Y$ takes more than two values, one can easily create examples showing this equivalence does not hold. 
\vspace{1em}

\noindent
{\bf{Theorem 1}}
\vspace{1em}

When both random variables $X$ and $Y$ take two distinct finite values as set up above, and
\begin{equation}E[XY] = E[X]E[Y],
\label{noc2}
\end{equation} 
then
\begin{equation}P(X:Y) = P(X)P(Y).
\label{ind2}
\end{equation}
\vspace{1em}

\noindent
{\bf{Proof}}
\vspace{1em}

Let us define all the relevant probabilities with three parameters using
relations (\ref{jp}), (\ref{ep}) and (\ref{sp}). We set
\begin{equation}
\alpha = p_{11},\quad u = p({x_1}), \quad v = p({y_1}).\nonumber
\end{equation}
Then, other relevant probabilities can be expressed as summarized in the following updated table.

\begin{figure}[h]
\begin{center}
\includegraphics[width=0.23\columnwidth]{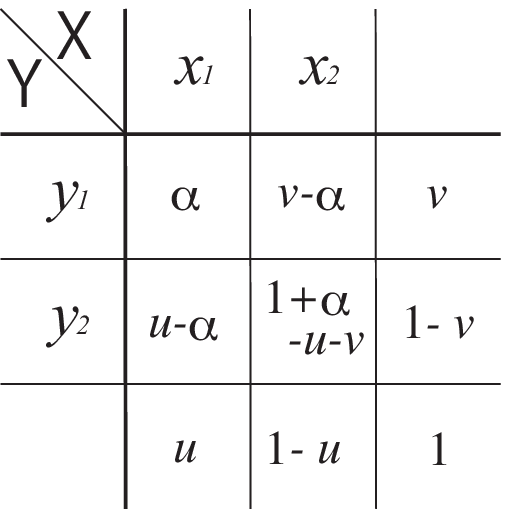}
\end{center}
\label{sind2}
\end{figure}

By definition of the expectation values, we have the following
\begin{eqnarray}
E[X] &=& u x_1 + (1-u) x_2, \nonumber\\
E[Y] &=& v y_1 + (1-v) y_2, \nonumber\\
E[XY] &=& \alpha {x_1}{y_1} +(u - \alpha){x_1}{y_2} + (v - \alpha){x_2}{y_1} + (1 - v - u + \alpha){x_2}{y_2}.\nonumber
\end{eqnarray}
The condition 
(\ref{noc2}) 
is now used together with above so that we obtain
\begin{eqnarray}
0 &=& E[XY] - E[X]E[Y] \nonumber\\
&=& \{ \alpha {x_1}{y_1} +(u - \alpha){x_1}{y_2} + (v - \alpha){x_2}{y_1} + (1 - v - u + \alpha){x_2}{y_2}\} \nonumber\\
&\ & \quad - \{u x_1 + (1-u) x_2 \}\{v y_1 + (1-v) y_2\} \nonumber\\
&=& (\alpha - u v)(x_1 - x_2)(y_1 - y_2).\nonumber
\end{eqnarray}

By the assumption that both of these stochastic variables take two distinct values ($x_1 \neq x_2,y_1 \neq y_2$), this leads to
\begin{equation}
\alpha - u v = p_{11} - p({x_1})p({y_1}) = 0.\nonumber
\end{equation}
From here, we can show that other joint probabilities can also be decomposed, leading to 
\begin{equation}P(X:Y) = P(X)P(Y). \quad Q.E.D.
\end{equation}

We note that above theorem is obtained even though we have eight parameters (four joint probabilities and $x_{1,2}, y_{1,2}$) with only two equations of constraints:(\ref{noc2}) and the relation that the sum of the four joint probabilities is equal to 1.

\section{Quantum case}

We now address the same question in quantum mechanics. In particular, we consider the analogous situation using
a pair of quantum particles, each taking two states. Such quantum particles are called qubits, and often used in the context of quantum information theory as a unit of information (e.g., \cite{wootters,schumacher,nielsen}). They are also used (e.g.,\cite{bell,clauser,leggett,ohira,pearle,popsecu}) in considering foundations of 
quantum mechanics and quantum measurement theories (e.g., \cite{araki,yanase,wheeler,allah}).

We consider the following. We have two quantum particles $A$ and $B$. 
The total normalized state vector $\ket{\psi}$ is given by

\begin{equation}
\ket{\psi}= \sum_{i,j} \gamma_{i,j}\ket{a_i}\otimes\ket{b_j}, \nonumber
\end{equation}
where $\ket{a}$ and $\ket{b}$ describe the state of particles $A$ and $B$, and $\gamma_{i,j}$ are quantum amplitudes given by complex scalars.

The two particle system is called separable when all $\gamma_{i,j}$ can be written as a product of two scalars.
\begin{equation}
\gamma_{i,j} = {\alpha_i}{\beta_j} \nonumber
\end{equation}
In other words, the system is separable when the total state vector can be decomposed as a product of each
normalized state of $A$ and $B$.
\begin{equation}
\ket{\psi} = \ket{\psi_A}\otimes\ket{\psi_B} \nonumber
\end{equation}
where
\begin{equation}
\ket{\psi_A}= \sum_{i} \alpha_i \ket{a_i}  \quad  \ket{\psi_B}= \sum_{j} \beta_j \ket{b_j}.\nonumber
\end{equation}

We make a natural correspondence between the notion of statistical independence with this separability of state vectors. The state vectors which are not separable are called entangled state vectors.

In order to consider expectation values of quantum observables,
we now set up two operators $\mathcal{X}$ and $\mathcal{Y}$ as
$\mathcal{X} = {\mathcal{Q}_A}\otimes{\bf{1}}_B$ and $\mathcal{Y} = {\bf{1}}_A\otimes\mathcal{R}_B$.
Here, 
${\mathcal{Q}_A}$ and ${\mathcal{R}_B}$ are quantum observable operators for $A$ and $B$ respectively, and ${\bf{1}}$
is the identity operator.

With the above set up, we first proceed to show that separability entails no-correlations.
In terms of equations, we want to show that if $\ket{\psi}$ is separable,
\begin{equation}
\bra{\psi}\mathcal{X}\mathcal{Y}\ket{\psi}= \bra{\psi}\mathcal{X}\ket{\psi}\bra{\psi}\mathcal{Y}\ket{\psi}.
\label{nocorrelation}
\end{equation}
This follows immediately from the definition of separability as
\begin{eqnarray}
\bra{\psi}\mathcal{X}\mathcal{Y}\ket{\psi} &=& \bra{\psi_A}\otimes\bra{\psi_B}\mathcal{X}\mathcal{Y}\ket{\psi_A}\otimes\ket{\psi_B}\nonumber\\ 
&=& \bra{\psi_A}\mathcal{Q}\ket{\psi_A}\bra{\psi_B}\mathcal{R}\ket{\psi_B}
= \bra{\psi}\mathcal{X}\ket{\psi}\bra{\psi}\mathcal{Y}\ket{\psi}.\nonumber
\end{eqnarray}
We note in passing that validity of this property is not limited to two-state particles, but for general separable systems.

Now, we limit ourselves to two-state systems and examine whether analog of the Theorem 1 is
valid in the quantum case. In other words, we want to examine the following statement.
\vspace{1em}

If, in the above setting, particles $A$ and $B$ take two states (as in qubits) and the no-correlation condition (\ref{nocorrelation}) is satisfied, then the state vector is separable.
\vspace{1em}

It turns out that the above statement does not hold in quantum systems. We show this by explicitly constructing a counter-example. In order to do this, we use a non-separable (entangled) Bell state vector, and 
give observable operators which satisfy (\ref{nocorrelation}).

Bell state vector is a representative example of entangled state vectors. We use the following Bell state for our two-particle, two-state system.
\begin{equation}
\ket{\phi}= {1\over\sqrt{2}}(\ket{a_1}\otimes\ket{b_1} + \ket{a_2}\otimes\ket{b_2}), \nonumber
\end{equation}
where $\ket{a_{(1,2)}}$ are the orthonormal states of the particle $A$, and similarly $\ket{b_{(1,2)}}$ for the particle $B$. This state is clearly not separable. 
We compute expectation values with respect to this Bell state vector. 

The expectation value for the product of the observable operators is given as
\begin{eqnarray}
\bra{\phi}\mathcal{X}\mathcal{Y}\ket{\phi} &=& {1\over 2} (\bra{a_1}\otimes\bra{b_1} + \bra{a_2}\otimes\bra{b_2})({\mathcal{Q}_A}\otimes{\bf{1}}_B)({\bf{1}}_A\otimes\mathcal{R}_B)(\ket{a_1}\otimes\ket{b_1} + \ket{a_2}\otimes\ket{b_2})\nonumber\\
&=&{1\over 2}(q_{11}r_{11} + q_{12}r_{12} + q_{21}r_{21} + q_{22}r_{22}), 
\label{leftcorr}
\end{eqnarray}
where $q_{ij} = \bra{a_i}\mathcal{Q}_A\ket{a_j}$ and $r_{ij} = \bra{b_i}\mathcal{R}_B\ket{b_j}$.

On the other hand, the product of expectation values for the two particles can be computed as follows.
\begin{eqnarray}
\bra{\phi}\mathcal{X}\ket{\phi}\bra{\phi}\mathcal{Y}\ket{\phi}
&=& {1\over 4} (\bra{a_1}\otimes\bra{b_1} + \bra{a_2}\otimes\bra{b_2})
({\mathcal{Q}_A}\otimes{\bf{1}}_B)
(\ket{a_1}\otimes\ket{b_1} + \ket{a_2}\otimes\ket{b_2})\nonumber\\
&\times &(\bra{a_1}\otimes\bra{b_1} + \bra{a_2}\otimes\bra{b_2})({\bf{1}}_A\otimes\mathcal{R}_B)(\ket{a_1}\otimes\ket{b_1} + \ket{a_2}\otimes\ket{b_2})\nonumber\\
&=& {1\over 4}((q_{11} + q_{22})(r_{11} + r_{22}))\nonumber\\
&=&{1\over 4}(q_{11}r_{11} + q_{22}r_{11} + q_{11}r_{22} + q_{22}r_{22}).
\label{rightcorr}
\end{eqnarray}

The condition for no-correlation is satisfied when (\ref{leftcorr}) is equal to (\ref{rightcorr}). We can achieve this by assigning various concrete values for the matrix elements $q_{ij}, r_{ij}$ of
${\mathcal{Q}_A}$ and ${\mathcal{R}_B}$ in the basis of $\ket{a_{(1,2)}}$ and $\ket{b_{(1,2)}}$ respectively. One example is shown below.
\begin{equation}
{\mathcal{Q}_A}=
\begin{bmatrix}
3 & 1 \\
1 & 1
\end{bmatrix}
,\quad
{\mathcal{R}_B}=
\begin{bmatrix}
1 & 1 \\
1 & 3
\end{bmatrix}
\end{equation}

The above observation means that no-correlation does not necessarily leads to separability for this two-particle two-state quantum system.
Thus, we have shown that Theorem 1, which holds for classical probabilities, does not carry over to analogous quantum systems.

\section{Conclusion}

We have considered a relation between statistical independence and correlation both in classical and quantum theories. In particular, we noted that for two-state two-variable systems, the classical probability show equivalence between the statistical independence and no-correlation. However, this equivalence in analogous quantum systems does not hold. Our observation provides yet another example of differences between classical and quantum theories.

\section*{Acknowledgment}

The author would like to thank Philip M. Pearle, Professor Emeritus of Hamilton College, for his comments and encouragements.
This work was supported by funding from Ohagi Hospital, Hashimoto, Wakayama, Japan, and by Grant-in-Aid for Scientific Research from Japan Society for the Promotion of Science No.16H01175, No.16H03360, No.18H04443.


%

\end{document}